\documentclass[10pt,a4paper]{article}
\usepackage[a4paper,left=1.5cm,right=1.5cm,top=2cm,bottom=2cm]{geometry}
\usepackage{cite}
\usepackage{amsmath,amssymb,amsfonts}
\usepackage{algorithmic}
\usepackage{iftex}
\ifPDFTeX
  \usepackage{graphicx}
\else
  \usepackage[dvipdfmx]{graphicx}
  
\fi
\usepackage{textcomp}
\usepackage{xcolor}
\usepackage{xurl}
\ifPDFTeX
  \usepackage[hidelinks]{hyperref}
\else
  \usepackage[dvipdfmx,hidelinks]{hyperref}
\fi
\usepackage{tikz}
\usetikzlibrary{tikzmark}
\usetikzlibrary{calc}
\usepackage{listings}
\usepackage{bm}
\usepackage{etoolbox}
\usepackage{abstract}
\usepackage{authblk}

\AtBeginEnvironment{table}{\footnotesize}
\AtBeginEnvironment{tabular}{\setlength{\tabcolsep}{2pt}}

\newcommand{\bracestart}[1]{\tikzmark{#1-start}}
\newcommand{\braceend}[1]{\tikzmark{#1-end}}
\newcommand{\drawlogannotations}[2]{%
  \begin{tikzpicture}[overlay,remember picture]
    \foreach \mark/\num in {#1} {
      \node[circle,draw,inner sep=1pt,minimum size=14pt] at ([xshift=1.6cm,yshift=0.5ex]pic cs:\mark) {\footnotesize \num};
      \draw ([xshift=0.05cm,yshift=0.5ex]pic cs:\mark) -- ([xshift=1.3cm,yshift=0.5ex]pic cs:\mark);
    }
    \foreach \mark/\num in {#2} {
      \draw ([xshift=0.6cm,yshift=0.5ex]pic cs:\mark-start) -- ++(0.15,0)
        -- ([xshift=0.75cm,yshift=0.5ex]pic cs:\mark-end)
        -- ++(-0.15,0);
      \node[circle,draw,inner sep=1pt,minimum size=14pt] at ([xshift=1.6cm,yshift=0.5ex]$(pic cs:\mark-start)!0.5!(pic cs:\mark-end)$) {\footnotesize \num};
      \draw ([xshift=0.75cm,yshift=0.5ex]$(pic cs:\mark-start)!0.5!(pic cs:\mark-end)$) -- ([xshift=1.3cm,yshift=0.5ex]$(pic cs:\mark-start)!0.5!(pic cs:\mark-end)$);
    }
  \end{tikzpicture}
}

\newcommand{\ctext}[1]{\leavevmode\raise0.2ex\hbox{\textcircled{\scriptsize{#1}}}}

\newcommand{\PARstart}[2]{#1#2}
\newcommand{\EOD}{}

\newenvironment{keywords}{%
  \par\smallskip\noindent\textbf{Keywords: }\ignorespaces
}{\par\medskip}

\def\BibTeX{{\rm B\kern-.05em{\sc i\kern-.025em b}\kern-.08em
    T\kern-.1667em\lower.7ex\hbox{E}\kern-.125emX}}

\title{Violet: Enabling Full Virtualization for M-mode RTOS on RISC-V}
\author[1]{Taro Kito}
\author[2]{Ryosuke Yamamoto}
\author[2]{Keisuke Horii}
\author[2]{Hiroki Masuda}
\author[1]{Koichi Mouri}
\affil[1]{Ritsumeikan University, Ibaraki, Osaka 567--8570, Japan
  (e-mail: \nolinkurl{tkito@asl.cs.ritsumei.ac.jp}, \nolinkurl{mouri@asl.cs.ritsumei.ac.jp})}
\affil[2]{Information Technology R\&D Center, Mitsubishi Electric Corporation, Kamakura, Kanagawa 247--8501, Japan
  \mbox{(e-mail: \nolinkurl{Yamamoto.Ryosuke@ce.MitsubishiElectric.co.jp}}, \nolinkurl{Horii.Keisuke@bc.MitsubishiElectric.co.jp}, \nolinkurl{Masuda.Hiroki@dc.MitsubishiElectric.co.jp})}
\date{}

\makeatletter
\patchcmd{\@maketitle}{\null}{%
  \null
  {\footnotesize\noindent
  This work has been submitted to the IEEE for possible publication.
  Copyright may be transferred without notice, after which this version may no longer be accessible.\par}%
}{}{\PackageError{violet}{Could not insert the preprint notice}{}}
\makeatother

\begin{document}
\twocolumn[
\begin{@twocolumnfalse}
\maketitle
\vspace{-3em}
{\centering\footnotesize Corresponding author: Taro Kito (e-mail: \nolinkurl{tkito@asl.cs.ritsumei.ac.jp})\par}

\begin{abstract}
In embedded systems, complex configurations may be required, such as the simultaneous execution of a real-time operating system (RTOS) and a general-purpose operating system (GPOS), or the operation of multiple RTOS instances. Embedded system hypervisors have been studied and developed to meet these requirements for architectures like ARM and x86. RISC-V is experiencing growing adoption in embedded systems and faces similar needs. However, RISC-V's virtualization support targets only U-mode (where applications run) and S-mode (where general-purpose OSs run) as virtualization levels. The M-mode, where RTOSs like FreeRTOS or Zephyr run, is excluded from virtualization. This means that, similar to architectures like ARM, running an RTOS on a Virtual Machine (VM) using methods based on virtualization support features is impossible. Therefore, this paper proposes the Violet hypervisor. Violet combines RISC-V's virtualization features with software-based emulation, enabling the execution of unmodified M-mode RTOSs. Evaluation verified the validity of the M-mode emulation functionality using RISC-V architecture tests. Furthermore, this was implemented on the SiFive HiFive Premier P550 hardware, demonstrating that existing RTOSs can run on Violet's VM and that coexistence with GPOSs like Linux is also possible. The performance evaluation also quantified the overhead introduced by M-mode emulation on M-mode CSR accesses, timer interrupt latency, and context switching.
\end{abstract}

\vspace{1\baselineskip}
\end{@twocolumnfalse}
]

\section{Introduction} \label{section:intro}

\PARstart{I}{n} embedded systems, complex configurations may be required, such as the simultaneous execution of a real-time operating system (RTOS) and a general-purpose operating system (GPOS), or the operation of multiple RTOS instances.
An RTOS is typically used for tasks requiring real-time performance, while a general-purpose OS like Linux is used for tasks such as GUI or network operations.
Embedded hypervisors have been extensively studied and developed \cite{martins_et_al:OASIcs.NG-RES.2020.3,xvisor,10.1145/3313808.3313816,6871203} to meet this demand for simultaneously executing multiple OSs while managing and sharing such hardware resources.
These embedded hypervisors leverage processor virtualization features, such as those provided by ARM, to reduce virtualization overhead when running guest OSs.
Furthermore, since they can run without modifying the guest OS, they offer high portability.

RISC-V is one of the processor architectures experiencing growing adoption in embedded and domain-specific systems\cite{riscv_annual_report_2025}.
Embedded systems using RISC-V also have similar needs for concurrently executing multiple OSs and running existing guest OSs without modification, and hypervisors effectively address them.
However, running an RTOS as a guest OS on RISC-V is challenging. RISC-V has three privilege levels: U-mode, S-mode, and M-mode. However, RISC-V's virtualization support targets only the U-mode, where applications run, and the S-mode, where general-purpose OSs run. The M-mode where RTOSs like FreeRTOS\cite{freertos} or Zephyr\cite{zephyr} run, is not a target for virtualization.
Therefore, existing M-mode RTOS binaries cannot be run on a VM without modification by using only hardware virtualization support features.

Based on the above background, this paper sets the following Research Questions (RQs):

\begin{itemize}
	\item RQ1: Can a hypervisor be realized to run an existing M-mode RTOS on RISC-V without modification?
	\item RQ2: Does the mechanism used by the hypervisor proposed in RQ1 to run existing M-mode RTOSs comply with the RISC-V architecture?
	\item RQ3: Can the hypervisor proposed in RQ1 enable simultaneous execution of an M-mode RTOS and a GPOS on real RISC-V hardware?
	\item RQ4: What performance cost is incurred when running an existing M-mode RTOS on the hypervisor proposed in RQ1?
\end{itemize}

To address these RQs, a method for executing an RTOS operating in M-mode on a hypervisor was previously investigated, and a hypervisor running on QEMU was implemented\cite{riscv-summit}. Building upon these results, this paper proposes Violet\cite{violet}, an embedded hypervisor for RISC-V that enables RTOS execution on a VM without modification, allowing multiple OSs to run concurrently. Furthermore, the proposed hypervisor was implemented on the SiFive HiFive Premier P550\cite{sifive-hifive-premier-p550} development board, which features a RISC-V core, and its operation was verified.

The contributions of this paper are as follows:
\begin{itemize}
	\item Proposal of the Violet hypervisor by combining RISC-V's virtualization support features with software-based M-mode processing emulation, enabling existing M-mode RTOSs to run without modification.
	\item Implementation of Violet on actual hardware, demonstrating concurrent execution of an M-mode RTOS (FreeRTOS) and a GPOS (Linux), while also evaluating the effectiveness of Violet's M-mode emulation functionality.
	\item Quantitative evaluation of the performance overhead introduced by M-mode emulation on CSR accesses, interrupt response, and context switching.
\end{itemize}

This paper describes the background and challenges in Section \ref{section:riscv_arch}, presents the proposed hypervisor Violet in Section \ref{section:violet}, and evaluates Violet in Section \ref{section:evaluation}. Section \ref{section:related} covers related works, and Section \ref{section:conclusion} concludes the paper.

\section{Background and Challenge} \label{section:riscv_arch}

\subsection{Overview of RISC-V}

RISC-V is an open Instruction Set Architecture (ISA) developed at the University of California, Berkeley~\cite{riscv-isa-unpriv,riscv-isa-man}.
It is license-free, allowing anyone to freely use, implement, and extend it.
The RISC-V instruction set has a modular structure consisting of a base integer instruction set (RV32I for 32-bit or RV64I for 64-bit) combined with optional extensions such as the multiplication extension (M), atomic extension (A), floating-point extensions (F/D), and compressed instruction extension (C).

RISC-V has three distinct privilege levels\cite{riscv-isa-man}: M-mode, S-mode, and U-mode.
M-mode represents Machine mode, S-mode represents Supervisor mode, and U-mode represents User mode.
M-mode has the highest privilege, S-mode has the next highest privilege, and U-mode has the lowest privilege.
Depending on the privilege level, the executable privileged instructions, accessible Control and Status Registers (CSRs), and memory regions are restricted.

\subsection{Control and Status Registers (CSR\lowercase{s})} \label{section:riscv_arch_csr}

CSRs are registers used to indicate or control the state of the CPU.
Table \ref{tab:riscv_csrs} shows the major RISC-V CSRs, their descriptions, and the privilege levels at which they are accessible.
CSRs accessible only in M-mode have prefixes starting with "m", while those accessible in S-mode or higher have prefixes beginning with "s".
\texttt{mstatus} and \texttt{sstatus} are registers that control the CPU state, while \texttt{mepc} and \texttt{sepc} are registers that store the PC value at the time of an exception.
\texttt{mtvec} and \texttt{stvec} are registers specifying trap vector addresses, while \texttt{mcause} and \texttt{scause} indicate exception/interrupt causes. The registers \texttt{mie} and \texttt{sie} are interrupt enable registers, and \texttt{mip} and \texttt{sip} are interrupt pending registers.
These CSRs perform the same functions but are accessible at different privilege levels.
Furthermore, \texttt{mstatus}, \texttt{mie}, and \texttt{mip} are supersets of \texttt{sstatus}, \texttt{sie}, and \texttt{sip}, and also contain the bit fields of \texttt{sstatus}, \texttt{sie}, and \texttt{sip}.
Therefore, by manipulating \texttt{mstatus}, \texttt{mie}, and \texttt{mip}, it is possible to control interrupts in S-mode.
Conversely, manipulating \texttt{sstatus}, \texttt{sie}, and \texttt{sip} does not allow control of interrupts in M-mode.

\subsection{Interrupt Controllers}\label{section:riscv_arch_int_controller}

Many RISC-V processors incorporate the Core-Local Interruptor (CLINT) and the Platform-Level Interrupt Controller (PLIC) as their interrupt controllers.

CLINT is an interrupt controller that manages software interrupts and timer interrupts in RISC-V\cite{clint}.
CLINT is locally connected to each CPU core and contains registers for raising software and timer interrupts.
These are accessible via Memory-Mapped I/O (MMIO).
CLINT is often managed by firmware operating in M-mode, such as OpenSBI\cite{opensbi}, so an OS operating in S-mode cannot directly manipulate CLINT.
Therefore, the OS operates CLINT via the Supervisor Binary Interface (SBI). The SBI is described in Section \ref{section:sbi}.

PLIC \cite{plic} is an interrupt controller that manages external interrupts in RISC-V.
Registers exist to enable/disable interrupts, set priorities, and check pending interrupt status per CPU core or privilege level, accessed via MMIO.
Using PLIC enables interrupt distribution among multiple CPU cores and interrupt control per privilege level.

\subsection{Supervisor Binary Interface (SBI)} \label{section:sbi}

SBI \cite{riscv-sbi} is a specification that abstracts platform-specific functionality in RISC-V and provides an interface between the OS and firmware. Through SBI, the OS can operate hardware resources such as interrupts and timers in a hardware-independent manner. SBI is implemented by firmware running in M-mode, and an OS running in S-mode sends requests to the firmware via SBI.

\begin{figure}[t]
	\centering
	\includegraphics[width=7cm,keepaspectratio]{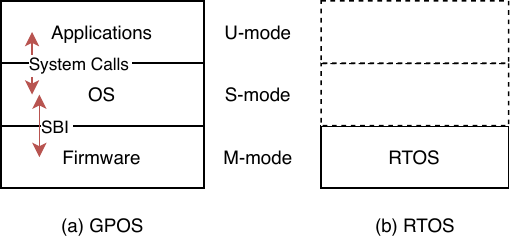}
	\caption{Privilege levels and software execution for GPOS and RTOS.}
	\label{fig:priv_software}
\end{figure}

\begin{table}[tb]
	\centering
	\caption{CSRs (Control and Status Registers) and their accessible privilege levels.}
	\label{tab:riscv_csrs}
	\resizebox{\linewidth}{!}{%
	\begin{tabular}{c|l|l}
		\hline
		\hline
		CSRs     & Description 																		 & \multicolumn{1}{c}{\begin{tabular}[c]{@{}l@{}}Accessible \\ privilege levels\end{tabular}}  \\
		\hline
		\texttt{mstatus}  & Register to control CPU status, & M     \\
		\texttt{sstatus}  & such as interrupt enable.																						 		 & M, S  \\
		\hline
		\texttt{mepc}     & Register indicating the PC value  & M      \\
		\texttt{sepc}     & when an exception occurs.															  								 & M, S   \\
		\hline
		\texttt{mtvec}    & Register that specifies  & M      \\
		\texttt{stvec}    & the address of the trap vector. 																							 & M, S   \\
		\hline
		\texttt{mcause}   & Register indicating  & M      \\
		\texttt{scause}   & the cause of the exception/interrupt.                                           		 & M, S   \\
		\hline
		\texttt{mie}      & Interrupt enable register. 												 & M      \\
		\texttt{sie}      & 																					  		 & M, S   \\
		\hline
		\texttt{mip}      & Interrupt pending register. 										 & M  		 \\
		\texttt{sip}      & 																					  		 & M, S   \\
		\hline
	\end{tabular}%
	}
\end{table}

\subsection{Execution Privilege Levels of GPOS and RTOS}\label{section:riscv_rtos_priv}

Figure \ref{fig:priv_software}(a) shows the relationship between privilege levels and the software that operates when executing a GPOS, and Figure \ref{fig:priv_software}(b) shows the corresponding relationship for an RTOS.
As shown in Figure \ref{fig:priv_software}(a), when executing a GPOS, firmware such as OpenSBI\cite{opensbi} operates in M-mode, a GPOS such as Linux\cite{linux} operates in S-mode, and applications operate in U-mode.
U-mode applications invoke system calls to transition to S-mode and request processing from the OS.
The S-mode OS uses SBI to request hardware-specific operations from the M-mode firmware.
In contrast, as shown in Figure \ref{fig:priv_software}(b), RTOSs such as FreeRTOS and Zephyr run in the highest privilege level, M-mode.
Therefore, an RTOS directly accesses M-mode CSRs, interrupt controllers, and other hardware resources, and typically does not use SBI.

\subsection{Hardware Virtualization Support: Hypervisor Extension} \label{section:riscv_hyp}

\begin{figure}[t]
	\centering
	\includegraphics[width=6cm,keepaspectratio]{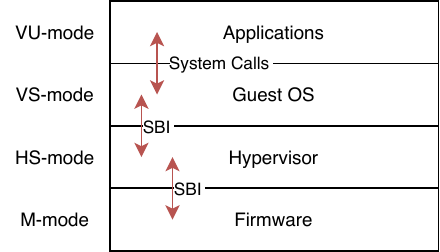}
	\caption{Privilege levels and software execution when Hypervisor Extension is enabled.}
	\label{fig:hyp_priv_software}
\end{figure}

The RISC-V privileged architecture specification defines the Hypervisor Extension as hardware support for hypervisor implementation\cite{riscv-isa-man}.
The Hypervisor Extension adds new privilege levels for hypervisors and VMs, along with corresponding CSRs and hypervisor instructions.
Figure \ref{fig:hyp_priv_software} illustrates the relationship between privilege levels and the software that operates at each level when the Hypervisor Extension is enabled.
Furthermore, since the hypervisor appears to guest OSs as privileged software equivalent to firmware, it must provide SBI. Guest OSs use SBI to configure timers and control interrupts between cores.
The hypervisor virtualizes these SBI requests when required.

\subsubsection{HS-mode}

HS-mode is the privileged level at which the hypervisor operates.
HS-mode replaces S-mode when Hypervisor Extension is enabled. HS-mode adds CSR and hypervisor instructions to S-mode.

Registers for controlling the two-stage address translation of the Memory Management Unit (MMU), used to isolate and protect memory spaces between guest OSs and between guest OSs and the hypervisor, as well as registers for managing the VM state, are added.

\subsubsection{VS-mode and VU-mode}
VS-mode is the privilege level at which the guest OS operates on the guest VM.
VS-mode is a virtualized privilege level of S-mode, providing the same set of CSRs and privileged instructions as S-mode.
However, in practice, the hypervisor provides and manages a dedicated set of CSRs for VS-mode.
When a guest OS attempts to access S-mode CSRs in VS-mode, the hardware translates this access to VS-mode CSRs.
Table \ref{tab:riscv_hyp_vs_csrs} shows this mapping.
For example, an attempt to access \texttt{sstatus} is translated into an access to \texttt{vsstatus}.
Therefore, even if the guest OS executes S-mode privileged instructions or references CSRs, no exception occurs, and no hypervisor trap is required.

\begin{table}[tb]
	\centering
	\caption{VS-mode CSRs and their corresponding S-mode CSRs.}
	\label{tab:riscv_hyp_vs_csrs}
	\begin{tabular}{c|c}
		\hline
		\hline
		VS-mode CSRs & S-mode CSRs \\
		\hline
		\texttt{vsepc} & \texttt{sepc} \\
		\hline
		\texttt{vsstatus} & \texttt{sstatus} \\
		\hline
		\texttt{vstvec} & \texttt{stvec} \\
		\hline
		\texttt{vscause} & \texttt{scause} \\
		\hline
		\texttt{vsie} & \texttt{sie} \\
		\hline
		\texttt{vsip} & \texttt{sip} \\
		\hline
		\end{tabular}
\end{table}

Furthermore, VU-mode is the privilege level at which user programs run on the guest OS.

\begin{figure*}[tb]
	\centering
	\includegraphics[width=17cm,keepaspectratio]{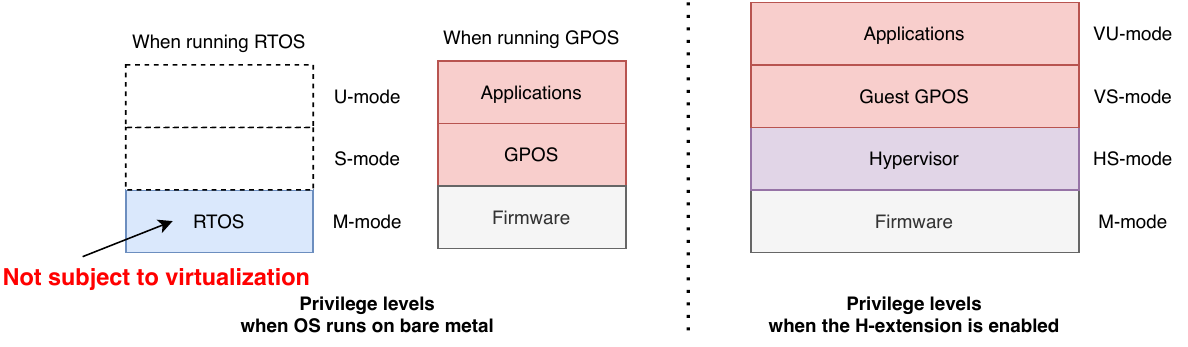}
	\caption{Privilege levels that are virtualized and those that are not.}
	\label{fig:rtos_priv-level}
\end{figure*}

\subsection{Challenge in Executing M-mode RTOS} \label{section:riscv_challange}

By utilizing the Hypervisor Extension described in Section \ref{section:riscv_hyp}, multiple hypervisors exist for RISC-V that execute an OS normally operating in S-mode within VS-mode,\cite{kvm,10.1145/1165389.945462,bhyve,bao}.
However, as shown in Figure \ref{fig:rtos_priv-level}, the Hypervisor Extension targets S-mode and U-mode for virtualization.
There is no mechanism to virtualize M-mode registers or privileged instructions.
Thus, because the Hypervisor Extension does not virtualize M-mode, the hypervisor above cannot run M-mode programs.

Therefore, as discussed in Section \ref{section:riscv_rtos_priv}, while RTOSs designed to run in M-mode exist, existing RISC-V hypervisors cannot support running multiple such RTOSs or simultaneously running a general-purpose OS and an RTOS.

\begin{figure*}[tb]
	\centering
	\includegraphics[width=15cm,keepaspectratio]{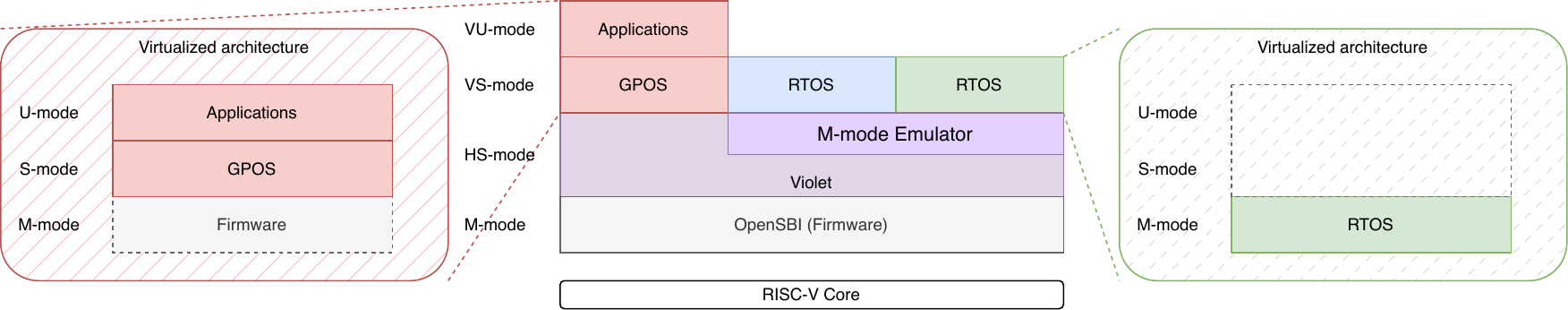}
	\caption{Overview of Violet.}
	\label{fig:violet_overview}
\end{figure*}

\section{Violet} \label{section:violet}

\subsection{Overview}

Violet\cite{violet} is an embedded hypervisor for RISC-V.
Figure \ref{fig:violet_overview} shows an overview of Violet.
In this architecture, Violet operates in HS-mode, while OpenSBI\cite{opensbi} operates as firmware in M-mode.
The following summarizes Violet's features for executing guest OSs in this configuration.

\begin{itemize}
	\item Capable of executing GPOS operating in S-mode
	\begin{itemize}
		\item GPOS operating in S-mode runs as a guest OS in VS-mode using the virtualization support features of the Hypervisor Extension.
	\end{itemize}
	\item Capable of executing RTOS operating in M-mode
	\begin{itemize}
		\item RTOS operating in M-mode runs as a guest OS in VS-mode by combining the virtualization support features of the Hypervisor Extension with the M-mode Emulator.
	\end{itemize}
	\item Capable of simultaneously executing multiple OSs
	\begin{itemize}
		\item By executing each guest OS as an independent VM in VS-mode, co-execution of GPOS and RTOS, or execution of multiple RTOSs, is possible.
	\end{itemize}
	\item Provides full virtualization
	\begin{itemize}
		\item Violet provides a virtualized hardware environment to guest OSs running in VS-mode, enabling execution on a VM without requiring modifications to the guest OS.
	\end{itemize}
	\item Device access employs pass-through mode
	\begin{itemize}
		\item Except for specific devices, Violet adopts a pass-through mode that allows guest OSs running in VS-mode to directly access assigned devices, minimizing virtualization overhead.
	\end{itemize}
\end{itemize}

\subsection{Execution of GPOS}

As usual, GPOS operating in S-mode runs on a VM using the Hypervisor Extension.
The guest GPOS runs in VS-mode, with applications operating in VU-mode.
As described in Section \ref{section:riscv_hyp}, S-mode privileged instructions are available in VS-mode,
and all accesses to S-mode CSRs are hardware-translated into accesses to VS-mode CSRs.
Therefore, software's virtualization of CPU features is unnecessary. Furthermore, the guest GPOS can access memory via its virtual and physical address spaces using two-stage address translation.

\subsubsection{SBI Virtualization}

The guest OS uses SBI to configure timers and control interrupts between cores, so Violet implements and provides SBI. The hypervisor virtualizes and processes requests via SBI, forwarding them to OpenSBI as necessary.
This makes it appear to the guest GPOS running in VS-mode that the firmware implementing SBI is running in M-mode.

\subsubsection{Device Pass-through}

The memory spaces of guest OSs, including device MMIO regions, are isolated by the two-stage address translation of the Hypervisor Extension.
For device access, except for the PLIC external interrupt controller, Violet maps the MMIO regions statically assigned to each guest OS into its guest physical address space, thereby enabling direct access to the assigned devices.
This mechanism controls whether a guest can access statically assigned address ranges; it does not control interference through shared hardware resources such as shared caches or memory bandwidth.
Such bandwidth or QoS control would require hardware that implements RISC-V QoS specifications such as Ssqosid or CBQRI\cite{riscv-ssqosid,riscv-cbqri}.
However, hardware implementations of these specifications are not currently available; therefore, strict interference isolation for shared hardware resources is outside the scope of this work.

\subsubsection{PLIC Virtualization}

PLIC can notify software running in M-mode and S-mode of external interrupts. However, it cannot notify guest operating systems operating in VS-mode of external interrupts.
To virtualize PLIC, interrupts intended for S-mode must be trapped each time and converted into interrupts for VS-mode for notification.
At this point, the PLIC state the guest OS assumes may differ from the actual PLIC state. Therefore, instead of allowing the guest OS to use the PLIC directly, Violet utilizes a Virtual PLIC (VPLIC).

\subsection{Execution of RTOS}

Existing RISC-V hypervisors cannot execute unmodified RTOSs designed to run in M-mode. Violet enables such RTOSs to run on a VM.
Violet combines the virtualization support features of the Hypervisor Extension with an M-mode Emulator to emulate M-mode within VS-mode. The RTOS operates within this emulation environment.
\subsection{M-mode Emulator}

\begin{figure}[tb]
	\centering
	\includegraphics[width=\columnwidth,keepaspectratio]{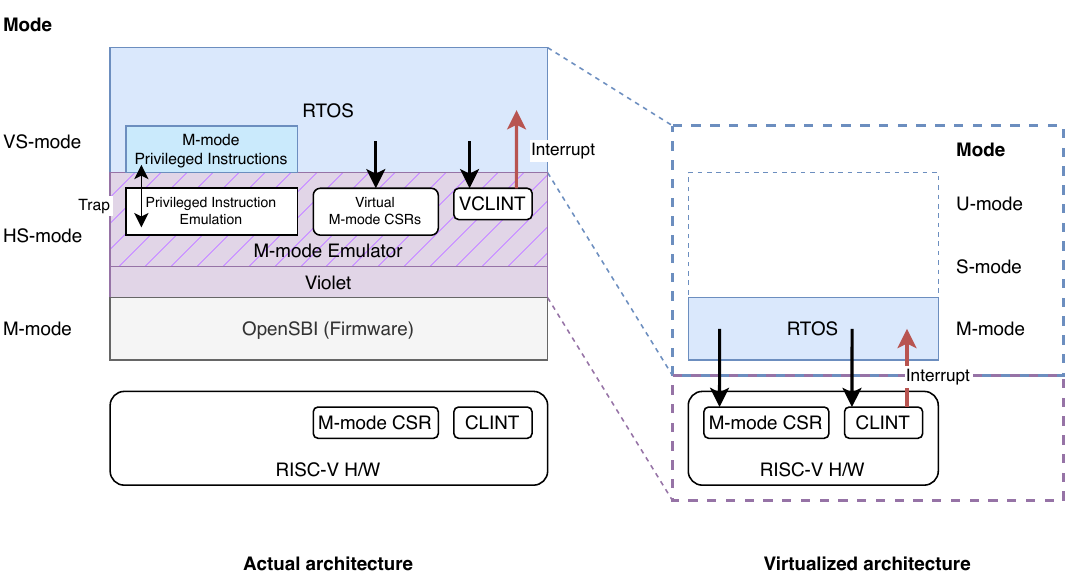}
	\caption{M-mode Emulator Overview.}
	\label{fig:violet_impl_vmmode}
\end{figure}

Figure \ref{fig:violet_impl_vmmode} shows an overview of the M-mode Emulator.
As discussed in Section \ref{section:riscv_challange}, RISC-V's Hypervisor Extension excludes M-mode from virtualization. Consequently, unlike other architectures, running an RTOS on a VM is impossible using virtualization support features.
Therefore, Violet provides an M-mode Emulator that runs the RTOS on a VM by emulating M-mode in software.
To virtualize M-mode, it provides the following features:

\begin{itemize}
	\item Emulation of privileged instructions
	\item Emulation of CSR accesses
	\item Emulation of interrupts and exceptions
	\item Emulation of CLINT
\end{itemize}

\subsubsection{Emulation of Privileged Instructions}

When executing RTOS or bare-metal programs on a VM using Violet's M-mode Emulator, these M-mode software programs run in VS-mode.
VS-mode is a virtualization of S-mode and operates at a lower privilege level than M-mode.
Therefore, executing M-mode privileged instructions in VS-mode is impossible and results in an Illegal Instruction exception.
The M-mode Emulator traps this exception and emulates the M-mode privileged instruction,
enabling the execution of M-mode privileged instructions within VS-mode.

\begin{figure*}[tb]
	\centering
	\begin{tabular}{cc}
		\includegraphics[width=0.47\linewidth,keepaspectratio]{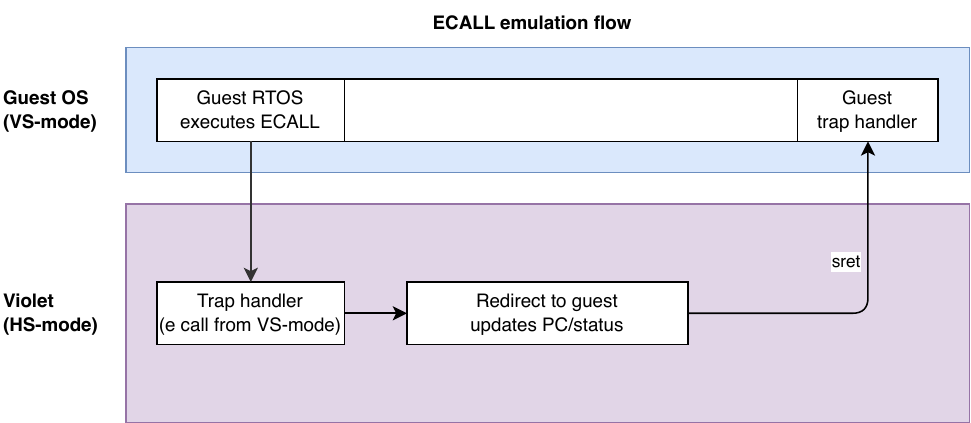} &
		\includegraphics[width=0.47\linewidth,keepaspectratio]{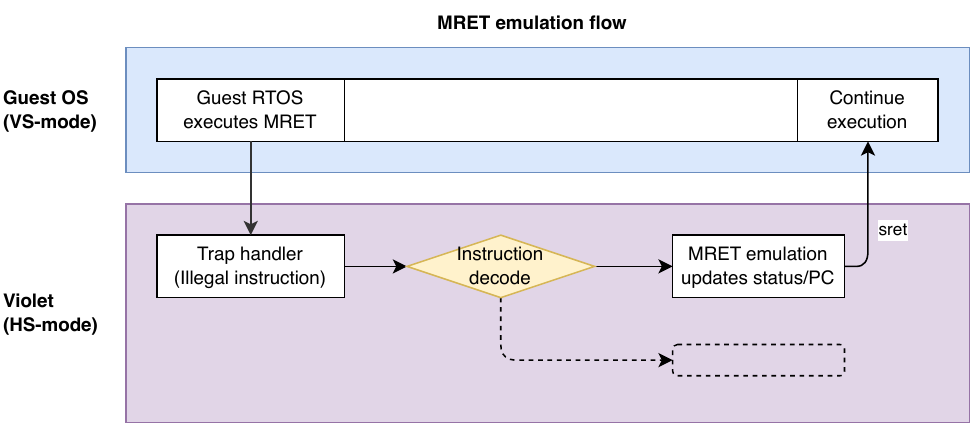} \\
		(a) \texttt{ECALL} emulation &
		(b) \texttt{MRET} emulation
	\end{tabular}
	\caption{Privileged Instructions emulation flows.}
	\label{fig:ecall_mret_flows}
\end{figure*}

Table \ref{tab:m-mode_emu} shows the M-mode privileged instructions requiring emulation.
\texttt{ECALL} is an instruction that performs an environment call. Normally, when \texttt{ECALL} executes in M-mode, an Environment Call from M-mode exception occurs, and the execution jumps to the M-mode trap handler.
However, when an RTOS executes \texttt{ECALL} on Violet's VM, an Environment Call from VS-mode exception occurs, causing a jump to Violet's trap handler in HS-mode.
Therefore, Violet must trap \texttt{ECALL} when executed in VS-mode and emulate it to behave identically when \texttt{ECALL} is executed in M-mode.
Specifically, Violet in HS-mode traps the Environment Call from VS-mode exception raised by executing \texttt{ECALL} in VS-mode.
It then invokes the guest RTOS trap handler in VS-mode to emulate \texttt{ECALL}.
Furthermore, since the exception type raised differs from the original, separate exception emulation is required. This is explained in Section \ref{section:trap-emulation}.

\texttt{MRET} is a privileged instruction that returns from M-mode to M/S/U-mode, primarily used for returning from trap handlers.
Since this instruction can only be executed in M-mode, executing \texttt{MRET} in VS-mode causes an Illegal Instruction exception.
The M-mode Emulator emulates \texttt{MRET} by trapping this exception and replacing it with an \texttt{SRET} instruction for execution.

\begin{table}[tb]
	\centering
	\caption{M-mode instructions that need to be emulated.}
	\label{tab:m-mode_emu}
	\begin{tabular}{c|l}
		\hline
		\hline
		Instruction & Description                          \\
		\hline
		ECALL       & Environment call                     \\
		            & raises an environment call exception \\
		            & (For system calls and SBI calls)     \\
		MRET        & Return from M-mode into M/S/U-mode   \\
		            & (Return from trap handler)           \\
		\hline
	\end{tabular}
\end{table}

Figure \ref{fig:ecall_mret_flows} shows the emulation flows for \texttt{ECALL} and \texttt{MRET}.
In Fig. \ref{fig:ecall_mret_flows}(a), when the guest RTOS executes \texttt{ECALL} in VS-mode, an Environment Call from VS-mode exception occurs and control transfers to Violet's trap handler in HS-mode.
Violet then sets the guest exception state so that the event appears as the exception handling that would occur after executing \texttt{ECALL} in M-mode, and resumes execution at the guest RTOS trap handler.
In Fig. \ref{fig:ecall_mret_flows}(b), when the guest RTOS executes \texttt{MRET} in VS-mode, an Illegal Instruction exception occurs and Violet catches it.
Violet restores the return destination and interrupt state from the virtualized M-mode state, and returns to VS-mode guest execution using \texttt{SRET}.

\subsubsection{Emulation of CSR accesses} \label{section:csr-emulation}

An Illegal Instruction exception occurs when access to the CSR in M-mode occurs in VS-mode. Implementing a virtual CSR within the M-mode Emulator and emulating access to the M-mode CSR enables access to the M-mode CSR in VS-mode.

The CSR that requires virtualization is classified into three categories (a) to (c) for explanation.

\begin{description}
	\item[(a) CSRs replaceable with VS-mode CSRs] \mbox{} \\
		These CSRs can be emulated by replacing accesses to them with accesses to VS-mode CSRs, since VS-mode also implements CSRs with equivalent functionality.
        For example, access to the mtvec register can be emulated by mapping it to the vstvec register.
        However, CSRs such as mstatus and mcause require value conversion during both write and read operations because the values stored in the corresponding VS-mode CSRs are not identical.
	\item[(b) PMP-related CSRs] \mbox{} \\
        These CSRs are related to Physical Memory Protection (PMP). M-mode software manipulates these CSRs to control memory access via PMP. However, since access to PMP-related CSRs is not possible from VS-mode or S-mode, Violet must emulate both these CSR accesses and the memory protection provided by PMP.
        Similarly, where Violet operates in HS-mode, it cannot manipulate PMP-related CSRs.
        Therefore, it cannot perform memory protection using PMP, but it can emulate it using paging.
	\item[(c) Read-only CSRs] \mbox{} \\
		Since these CSRs are read-only registers, Violet emulates access to them by returning arbitrary values.
\end{description}

In Violet's M-mode Emulator, the following registers from categories (a) and (c) were implemented as virtual registers: \texttt{mhartid}, \texttt{mstatus}, \texttt{mie}, \texttt{mip}, \texttt{mcause}, \texttt{mepc}, \texttt{mtvec}, \texttt{mideleg}, \texttt{medeleg}, \texttt{mtval}, \texttt{mscratch}, and \texttt{misa}.
Accessing these registers triggers an Illegal Instruction exception. Violet traps this exception, performs appropriate emulation for each register, and returns control to the guest OS.

\begin{figure}[tb]
	\centering
	\includegraphics[width=1\linewidth,keepaspectratio]{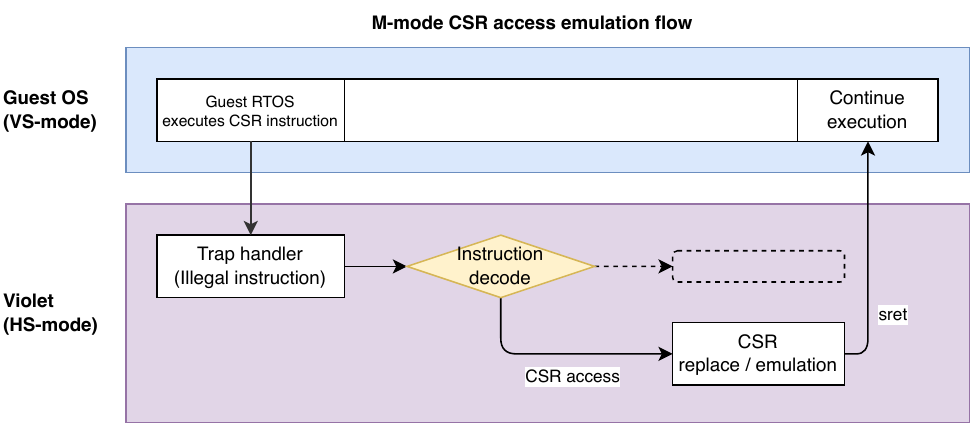}
	\caption{M-mode CSR access emulation flow.}
	\label{fig:csr_access_flow}
\end{figure}

\begin{figure}[t]
	\centering
	\includegraphics[width=\columnwidth,keepaspectratio]{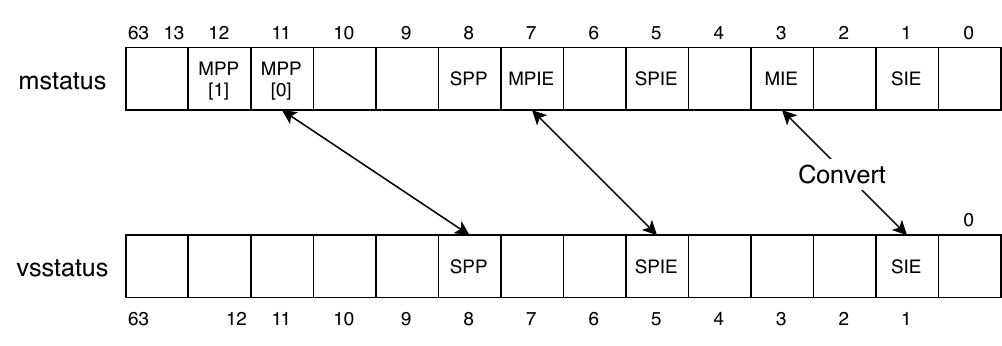}
	\caption{Conversion of values between mstatus and vsstatus.}
	\label{fig:violet_impl_status}
\end{figure}

\begin{figure}[t]
	\centering
	\includegraphics[width=\columnwidth,keepaspectratio]{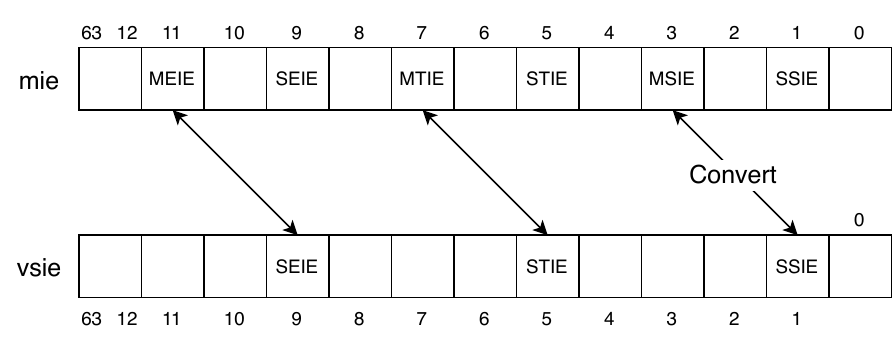}
	\caption{Conversion of values between mie and vsie.}
	\label{fig:violet_impl_ie}
\end{figure}

\begin{figure}[t]
	\centering
	\includegraphics[width=\columnwidth,keepaspectratio]{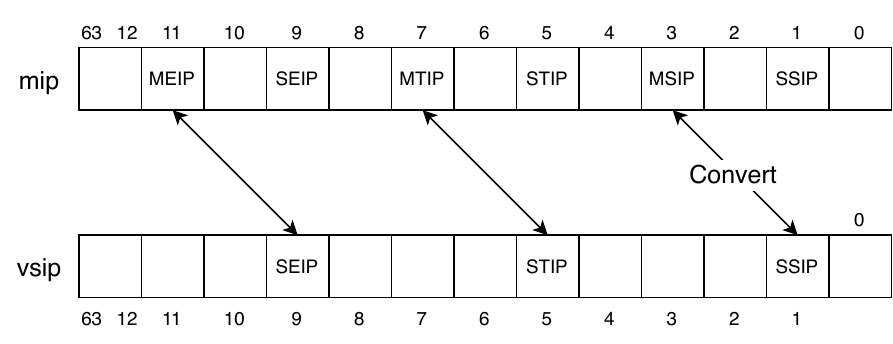}
	\caption{Conversion of values between mip and vsip.}
	\label{fig:violet_impl_ip}
\end{figure}

Figure \ref{fig:csr_access_flow} shows the emulation flow for M-mode CSR accesses.
When the guest RTOS accesses an M-mode CSR in VS-mode, an Illegal Instruction exception occurs and control transfers to Violet's trap handler.
Violet decodes the trapping instruction and determines both the target CSR and the access type.
If the target CSR is implemented as a virtual CSR, Violet reads or writes the virtual CSR inside the M-mode Emulator.
If the target CSR can be replaced with a VS-mode CSR, Violet converts the bit fields as needed and accesses the corresponding VS-mode CSR.
After the emulation, Violet advances the guest program counter and returns control to the guest RTOS in VS-mode.
The bit-field conversions used for interrupt-related CSRs are detailed in Figs. \ref{fig:violet_impl_status}--\ref{fig:violet_impl_ip}.

\begin{figure}[t]
	\centering
	\includegraphics[width=\columnwidth,keepaspectratio]{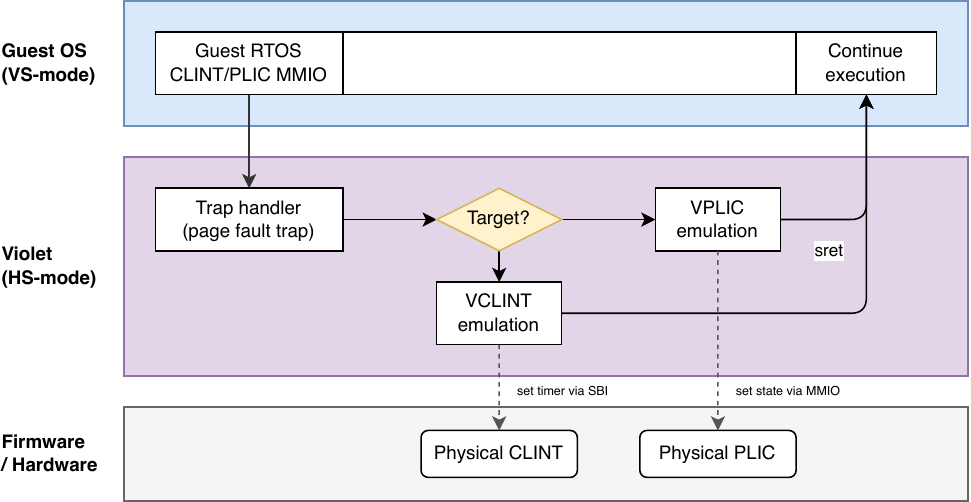}
	\caption{CLINT/PLIC MMIO emulation flow.}
	\label{fig:pic_mmio_emulation_flow}
\end{figure}

\begin{figure}[t]
	\centering
	\includegraphics[width=\columnwidth,keepaspectratio]{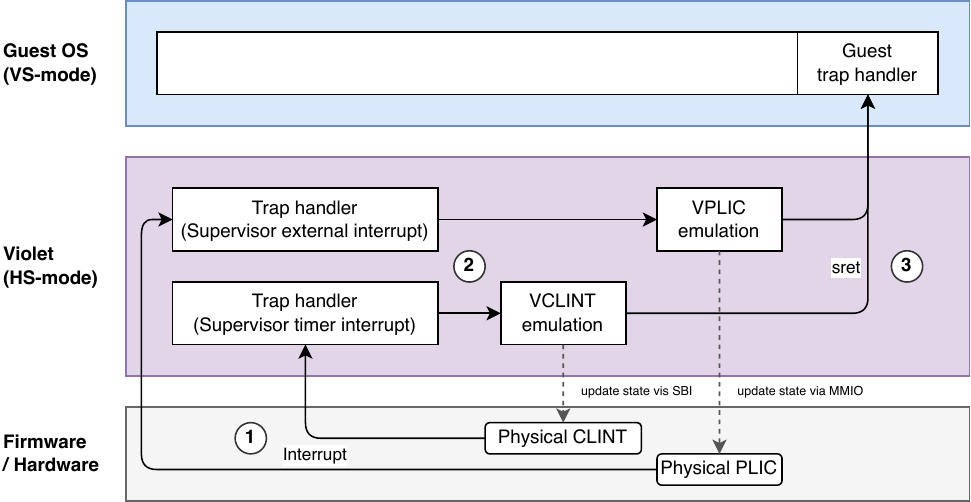}
	\caption{Interrupt injection flow.}
	\label{fig:interrupt_injection_flow}
\end{figure}

\subsubsection{Interrupt and Exception Emulation} \label{section:trap-emulation}

Table \ref{tab:traps} shows interrupts and exceptions for M-mode and interrupts and exceptions for S-mode.
RTOS and bare-metal programs designed to run in M-mode utilize interrupts and exceptions specific to M-mode.
Therefore, they cannot utilize these interrupts and exceptions when running in VS-mode.
Consequently, the M-mode Emulator must convert interrupts and exceptions intended for M-mode into those for VS-mode.

To convert interrupts and exceptions, Violet traps accesses to the \texttt{mstatus}, \texttt{mie}, and \texttt{mcause} registers.
It then hooks the read and write operations on these values to perform the conversion.
Emulation of the CSR (category (a) as described in \ref{section:csr-emulation}) enables this functionality.

\begin{table*}[t]
	\centering
	\caption{Conversion of cause values between guest-visible mcause and internal vscause.}
	\label{tab:cause_values}
	\small
	\begin{tabular}{l|l|l|l}
	\hline
	\hline
	Guest-visible M-mode cause & \texttt{mcause} value & Internal VS-mode cause & \texttt{vscause} value \\
	\hline
	\begin{tabular}[c]{@{}l@{}}Machine software interrupt\\(interrupt, no. 3)\end{tabular} & \hbox{0x8000000000000003} & \begin{tabular}[c]{@{}l@{}}Supervisor software interrupt\\(interrupt, no. 1)\end{tabular} & \hbox{0x8000000000000001} \\
	\hline
	\begin{tabular}[c]{@{}l@{}}Machine timer interrupt\\(interrupt, no. 7)\end{tabular} & \hbox{0x8000000000000007} & \begin{tabular}[c]{@{}l@{}}Supervisor timer interrupt\\(interrupt, no. 5)\end{tabular} & \hbox{0x8000000000000005} \\
	\hline
	\begin{tabular}[c]{@{}l@{}}Machine external interrupt\\(interrupt, no. 11)\end{tabular} & \hbox{0x800000000000000b} & \begin{tabular}[c]{@{}l@{}}Supervisor external interrupt\\(interrupt, no. 9)\end{tabular} & \hbox{0x8000000000000009} \\
	\hline
	\begin{tabular}[c]{@{}l@{}}Environment call from M-mode\\(exception, no. 11)\end{tabular} & \hbox{0x000000000000000b} & \begin{tabular}[c]{@{}l@{}}Environment call from VS-mode\\(exception, no. 9)\end{tabular} & \hbox{0x0000000000000009} \\
	\hline
	\end{tabular}
\end{table*}

\begin{table}[t]
	\centering
	\caption{Interrupts and exceptions for M-mode and VS-mode.}
	\label{tab:traps}
	\begin{tabular}{l|l}
		\hline
		\hline
		For M-mode              & For VS-mode                \\
		\hline
		Machine software        & Supervisor software        \\
		interrupt               & interrupt                  \\
		\hline
		Machine timer interrupt & Supervisor timer interrupt \\
		\hline
		Machine external        & Supervisor external        \\
		interrupt               & interrupt                  \\
		\hline
		Environment call        & Environment call           \\
		from M-mode             & from VS-mode               \\
		\hline
	\end{tabular}
\end{table}

The \texttt{mstatus} register contains the \texttt{mstatus.MIE} field, which sets interrupt enable/disable. Setting this to 1 globally enables interrupts for M-mode. However, since interrupts used on the VM are actually for VS-mode, it is necessary to enable interrupts for VS-mode, not M-mode.
Therefore, when the guest RTOS writes a value to \texttt{mstatus}, Violet traps this operation, converts the \texttt{mstatus} value, and writes it to \texttt{vsstatus}.
Figure \ref{fig:violet_impl_status} shows the mapping between the bit fields of \texttt{mstatus} to be converted and the bit fields of \texttt{vsstatus}.
For example, setting \texttt{mstatus.MIE} is converted to setting \texttt{vsstatus.SIE}. When \texttt{mstatus} is read, the value of \texttt{vsstatus} is converted to the corresponding bit field in \texttt{mstatus} and returned.

Similarly, the \texttt{mip} and \texttt{mie} registers are virtualized by mapping them to the corresponding bit fields in the vsip and vsie registers.
Since the \texttt{mip} register is used to insert interrupts for M-mode by type, and the \texttt{mie} register is used to enable or disable interrupts, these must be converted for VS-mode.
Figure \ref{fig:violet_impl_ie} and Figure \ref{fig:violet_impl_ip} show the bitfield mappings between \texttt{mie} and \texttt{vsie}, and between \texttt{mip} and \texttt{vsip}.
Read and write operations are performed by converting according to this mapping.

Figure \ref{fig:pic_mmio_emulation_flow} shows the MMIO access flow in CLINT/PLIC emulation.
When the guest RTOS accesses the MMIO region of CLINT or PLIC, a page fault occurs and control transfers to Violet in HS-mode.
Violet determines whether the access targets CLINT or PLIC, and dispatches the access to the VCLINT or VPLIC emulation path.
The emulation path updates the guest-visible CLINT/PLIC state and, if necessary, configures the physical CLINT/PLIC state through SBI or MMIO before resuming guest execution.
Figure \ref{fig:interrupt_injection_flow} shows the interrupt-injection flow.
When an interrupt arrives from the physical CLINT or PLIC, Violet's trap handler first receives the interrupt in HS-mode.
Violet then updates the VCLINT or VPLIC state, sets the corresponding VS-mode interrupt-pending state, and injects the interrupt into the guest RTOS.
Because the guest RTOS reads the state converted as an M-mode interrupt through CSR emulation, it can handle the interrupt as if it had been received in M-mode.

On the other hand, the \texttt{mcause} register stores the cause of the interrupt or exception. It is a register referenced within interrupt and exception handlers, used by software to identify the interrupt or exception that occurred.
Although the interrupts and exceptions that actually occur are intended for VS-mode, to make it appear as if an interrupt or exception intended for M-mode has occurred, it is necessary to convert the corresponding VS-mode interrupts and exceptions to M-mode interrupts and exceptions, as shown in Table \ref{tab:traps}.
Table \ref{tab:cause_values} shows the corresponding cause names, cause numbers, and actual values for \texttt{mcause} and \texttt{vscause}.

\subsubsection{CLINT Emulation}

In some cases, M-mode firmware, such as OpenSBI, restricts memory access to specific regions or memory-mapped devices.

Specifically, access to CLINT, which controls timers and inter-core interrupts, is restricted.
Therefore, Violet operating in HS-mode or a guest OS operating in VS-mode cannot access them.
However, since RTOSs, which inherently operate in M-mode, directly manipulate CLINT to control timer interrupts and inter-core interrupts,
Violet provides a virtualized CLINT, called Virtual CLINT (VCLINT), to RTOSs operating in VS-mode.

VCLINT implements the \texttt{mtime} register and the \texttt{mtimecmp} register to control timer interrupts, just like the actual CLINT.
When accessing CLINT from VS-mode, this access is replaced with access to VCLINT.
VCLINT operates the CLINT via SBI to set the timer when its registers are manipulated.

\section{Evaluation} \label{section:evaluation}

This section evaluates whether the M-mode Emulator provides an M-mode execution environment consistent with the RISC-V specification using RISC-V architecture tests.
It also evaluates whether the proposed hypervisor, Violet, can run an RTOS operating in M-mode and support a mixed environment of RTOS and GPOS.
It further evaluates the performance overhead introduced by the M-mode Emulator.

The evaluation items are as follows.

\begin{itemize}
	\item Functional evaluation
	\begin{itemize}
		\item Architecture-test-based evaluation of the M-mode Emulator
		\item Evaluation of FreeRTOS execution
		\item Evaluation of concurrent execution of Linux and FreeRTOS
	\end{itemize}
	\item Evaluation of performance overhead
\end{itemize}

Each evaluation item is described below. Table \ref{tab:test_env} shows the test environment and the relevant hardware specifications.

\begin{table}[t]
	\centering
	\caption{Test environment and relevant hardware specifications.}
	\label{tab:test_env}
	\begin{tabular}{p{0.25\columnwidth}|p{0.62\columnwidth}}
		\hline
		\hline
		Item       & Description                     \\
		\hline
		Board      & SiFive HiFive Premier P550\cite{sifive-hifive-premier-p550} \\
		SoC        & Eswin EIC7700X                  \\
		CPU        & Quad-core SiFive P550, 1.4 GHz  \\
		Memory     & 16 GB LPDDR5                    \\
		Hypervisor & Violet                          \\
		Guest OS   & Linux, FreeRTOS,                \\
		           & Test programs (riscv-arch-test) \\
		\hline
	\end{tabular}
\end{table}

\subsection{Functional Evaluation}

\subsubsection{Architecture-Test-Based Evaluation of the M-mode Emulator}

This subsubsection evaluates whether the M-mode execution environment provided by the M-mode Emulator behaves consistently with the RISC-V specification using RISCOF and riscv-arch-test.

RISCOF\cite{riscof} is a test framework for evaluating RISC-V specification compliance, and riscv-arch-test\cite{riscv-arch-test} is a test suite that provides test programs for the RISC-V specification. Based on the implemented RISC-V extensions and privilege levels described in a configuration file, RISCOF selects tests corresponding to the implementation under test from the riscv-arch-test pool and collects the results. In this evaluation, VioletVM is configured as a RISC-V implementation that implements only M-mode, and the selected tests are executed on Violet's VM.

\paragraph{Configuration}

Figure \ref{fig:riscof_configuration} shows the contents of the RISCOF configuration file (p550\_violetvm\_isa.yaml) used in this study. This configuration file defines the execution environment on a VM with Violet's M-mode Emulator enabled as a RISC-V implementation named VioletVM. The implemented RISC-V extension is specified as RV64IMAFDCZicsr\_Zifencei. This is the extension implemented in the RISC-V core of the SiFive HiFive Premier P550 used for testing, excluding the S-mode, U-mode, and Hypervisor Extension privilege levels, making it a RISC-V implementation with only M-mode.

\begin{figure}[t]
	\begin{lstlisting}
hart_ids: [0]
hart0:
	ISA: RV64IMAFDCZicsr_Zifencei
	physical_addr_sz: 51
	User_Spec_Version: '2.3'
	supported_xlen: [64]
\end{lstlisting}
	\caption{RISCOF Configuration for VioletVM.}
	\label{fig:riscof_configuration}
\end{figure}

With this configuration file, RISCOF executes the tests required for a RISC-V implementation that implements only M-mode on a Violet VM with the M-mode Emulator enabled.

\paragraph{Test Results}

Table \ref{tab:riscof_result} shows the RISCOF test results. Out of 674 tests conducted, 664 were PASSED, and 10 were FAILED.

The passed tests covered integer instructions, single-precision floating-point instructions, double-precision floating-point instructions, compressed instructions (excluding breakpoint), atomic instructions, \texttt{ECALL} instructions, and FENCE instructions. In particular, the \texttt{ECALL} instruction tests check exception handling caused by \texttt{ECALL} and the behavior of the exception-related CSRs updated during this process. Passing these tests indicates that the M-mode Emulator performs \texttt{ECALL} emulation, exception handling, and updates to the exception-related CSRs as expected. The 10 failed tests targeted misaligned load/store instructions and breakpoint instructions, as explained below.

The Misaligned Load/Store Instruction tests check whether Load Address Misaligned or Store/AMO Address Misaligned exceptions are raised for load/store instructions to unaligned addresses. However, for RISC-V implementations that do not raise these exceptions, these tests always FAIL and the results can be ignored\cite{riscv-arch-test}. Violet runs on OpenSBI, which executes in M-mode. OpenSBI traps these exceptions, emulates the memory accesses, and returns the results to Violet running in HS-mode. As a result, from the perspective of software running on the guest VM, VioletVM behaves as an implementation that does not raise exceptions for misaligned accesses. Therefore, these test results can be ignored.

The Breakpoint Instruction tests are for the breakpoint instruction. Since Violet does not virtualize the Breakpoint instruction, these tests always FAIL. Therefore, this result is expected and does not affect the functional evaluation.

For these reasons, the failed tests are either outside the scope of this evaluation or can be ignored. Since all other tests passed, the results confirm that VioletVM exhibits the expected behavior as an M-mode execution environment within the scope covered by riscv-arch-test.

\begin{table}[tb]
	\centering
	\caption{Test Results.}
	\label{tab:riscof_result}
	\begin{tabular}{lr}
		\hline
		\hline
		Test case                                   & Count \\
		\hline
		PASSED                                      & 664   \\
		\hline
		Integer Instructions                        &       \\
		Floating Point Instructions                 &       \\
		Double Floating Point Instructions          &       \\
		Atomic Instructions                         &       \\
		Compressed Instructions (except breakpoint) &       \\
		\texttt{ECALL} Instruction                  &       \\
		FENCE Instruction                           &       \\
		\hline
		FAILED                                      & 10    \\
		\hline
		Misaligned Load/Store Instruction           &       \\
		Breakpoint Instruction (include Compressed  &       \\
		Instructions)                               &       \\
		\hline
	\end{tabular}
\end{table}

\begin{figure}[t]
	\centering
	\includegraphics[height=7cm,keepaspectratio]{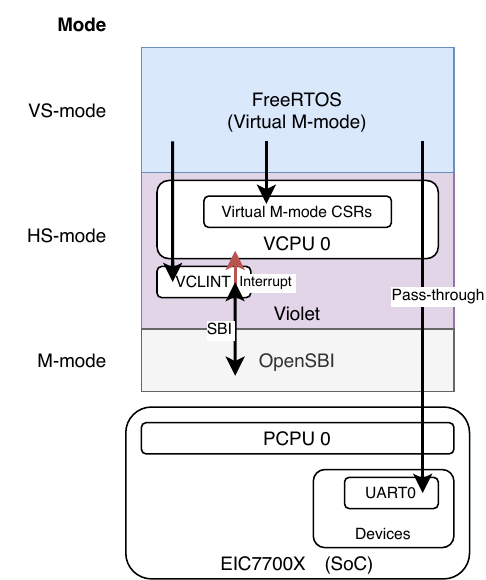}
	\caption{Overview of FreeRTOS execution on Violet.}
	\label{fig:freertos_on_violet}
\end{figure}

\subsubsection{Evaluation of FreeRTOS Execution}

This subsubsection evaluates whether FreeRTOS and its sample program \cite{freertos_p550} can execute on the virtual machine.

Figure \ref{fig:freertos_on_violet} shows the operating environment. VCPU0 is assigned to PCPU0, and FreeRTOS runs on it. In this case, the virtual M-mode realized by Violet's M-mode emulation function is used. Only CLINT, which handles timer interrupts, is virtualized, while other devices, such as UART, are passed through.

The FreeRTOS sample program to be executed displays "Hello FreeRTOS!" using UART at startup, and then outputs messages from the UART at regular intervals using timer interrupts. Successful execution of this sample program demonstrates that basic functions such as memory allocation to Violet's VM, trapping of interrupts and exceptions, and interrupt virtualization operate normally. Furthermore, since FreeRTOS is an RTOS that runs in M-mode, successful execution also indicates that Violet's M-mode Emulator, which performs M-mode CSR virtualization and privileged instruction emulation, is operating correctly.

As the execution result, Figure \ref{log:freertos} shows the execution log. This log was output using UART. \ctext{1} is the output from the firmware OpenSBI, \ctext{2} is the output from Violet, and the subsequent output is from FreeRTOS. \ctext{3} is the sample program output that periodically outputs messages using UART via timer interrupts. This shows that the M-mode Emulator, including M-mode CSR virtualization, privileged instruction emulation, and interrupt virtualization, is operating normally.

\begin{figure}[tb]
\begin{lstlisting}
Boot HART MIDELEG : 0x0000000000002666
Boot HART MEDELEG : 0x0000000000f00509 (*@\tikzmark{freertos_opensbi}@*)
Violet bootloader
Hello I'm Violet Hypervisor (*@\tikzmark{freertos_violet}@*)
Hello FreeRTOS!
0: Tx: Transfer1 (*@\bracestart{freertos_freertos}@*)
0: Rx: Blink1
0: Tx: Transfer2
0: Rx: Blink2    (*@\braceend{freertos_freertos}@*)
\end{lstlisting}
\caption{Log of FreeRTOS execution on Violet.}
\label{log:freertos}
\drawlogannotations{freertos_opensbi/1, freertos_violet/2}{freertos_freertos/3}
\end{figure}

\subsubsection{Evaluation of Concurrent Execution of Linux and FreeRTOS}

\begin{figure}[t]
  \centering
  \includegraphics[width=8cm,keepaspectratio]{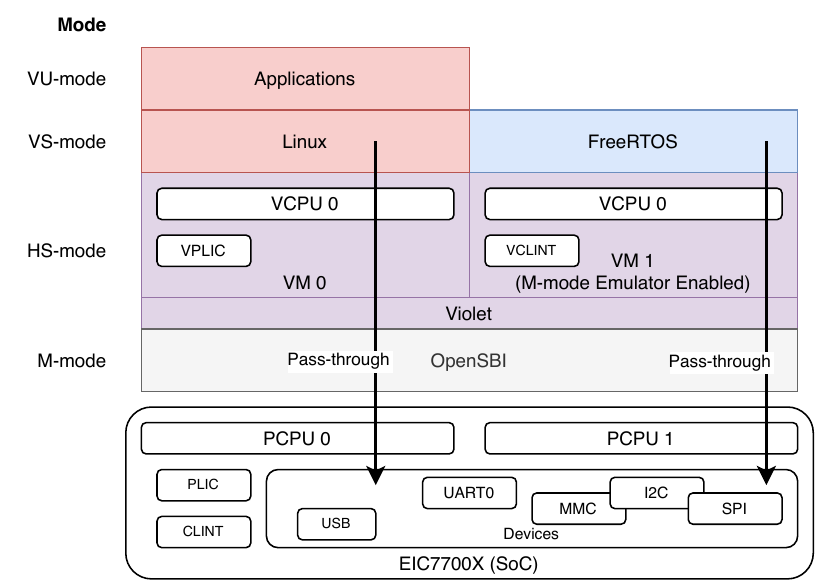}
  \caption{Overview of multiple VMs execution on Violet.}
  \label{fig:multi_vm}
\end{figure}

This subsubsection evaluates concurrent execution of Linux and FreeRTOS on Violet.
Figure \ref{fig:multi_vm} shows the operating environment when running multiple VMs.
Here, VCPU0 of VM0 is assigned to PCPU0, and Linux runs on it.
Also, VCPU0 of VM1 is assigned to PCPU1, and FreeRTOS runs on it.
For interrupt controller virtualization and device allocation, all devices except PLIC are passed through to the Linux guest of VM0 and assigned to it.
All devices, except CLINT, are passed through and assigned to the FreeRTOS guest in VM1.
UART0 is shared between VM0 and VM1 only for log output in this evaluation.

As the execution result, Figure \ref{log:multi_vm} shows the execution log. The logs until each guest OS starts up are omitted. This log was output using UART. \ctext{1} is the output of the FreeRTOS sample program on VM1. The rest is the Linux log on VM0. These results show that Linux and FreeRTOS run simultaneously on their respective VMs. The mixed log is due to the sharing of UART0.

\begin{figure}[t]
\begin{lstlisting}
0: Rx: Blink2
0: Tx: Transfer1
ubuntu@ubuntu:~$ uname -a
Linux ubuntu 6.6.21-9-premier #1 SMP PREEMPT_DYNAMIC
  Sat Nov  9 00:21:16 UTC 2024 riscv64 riscv64 riscv64 GNU/Linux
ubuntu@ubuntu:~$ 0: Rx: Blink1
0: Tx: Transfer2 (*@\bracestart{multi_freertos}@*)
0: Rx: Blink2
0: Tx: Transfer1
0: Rx: Blink1
0: Tx: Transfer2 (*@\braceend{multi_freertos}@*)
\end{lstlisting}
\caption{Log of multiple VMs execution.}
\label{log:multi_vm}
\drawlogannotations{}{multi_freertos/1}
\end{figure}

\subsection{Evaluation of Performance Overhead}

This subsection evaluates the performance overhead introduced by Violet's M-mode Emulator.
In Violet, M-mode CSR accesses and interrupt handling by the guest OS are trapped and handled by the M-mode Emulator.
Therefore, the overhead of M-mode CSR accesses can affect interrupt handling and RTOS scheduling behavior.
This evaluation measures M-mode CSR access latency, timer interrupt latency, and context-switch performance using the Thread-Metric Testsuite.

\subsubsection{M-mode CSR Access Latency}

The cycle count required for CSR read and write accesses was measured for the M-mode CSRs implemented by the M-mode Emulator.
Figure \ref{fig:csr_measurement_pseudocode} shows the measurement procedure as pseudocode.
In this measurement, the cycle counter was read before and after each CSR access instruction using \texttt{rdcycle}, and the difference was recorded as the CSR access latency.
This procedure was repeated 10000 times for each measured CSR read and write operation, and the first sample was excluded to reduce the effect of the initial cache miss.
Table \ref{tab:csr_overhead_all} shows the CSR access latency on VioletVM in cycles.

\begin{figure}[t]
\centering
\fbox{%
\begin{minipage}{0.94\linewidth}
\footnotesize
\begin{algorithmic}
	\FOR{each measured M-mode CSR}
		\FOR{each supported operation in \{\textit{read}, \textit{write}\}}
			\FOR{$i = 1$ to $10000$}
				\STATE $start \leftarrow \texttt{rdcycle}()$
				\STATE execute CSR operation
				\STATE $end \leftarrow \texttt{rdcycle}()$
				\STATE $sample[i] \leftarrow end - start$
			\ENDFOR
			\STATE discard the first sample
			\STATE compute min, max, average, and median
		\ENDFOR
	\ENDFOR
\end{algorithmic}
\end{minipage}%
}
	\caption{Pseudocode of the CSR access latency measurement.}
	\label{fig:csr_measurement_pseudocode}
\end{figure}

\begin{table}[t]
	\centering
	\caption{M-mode CSR access latency on VioletVM (cycles).}
	\label{tab:csr_overhead_all}
	\setlength{\tabcolsep}{4pt}
	\begin{tabular}{lrrrrrr}
		\hline
		\hline
		CSR & \begin{tabular}[c]{@{}c@{}}Read\\Min\end{tabular} & \begin{tabular}[c]{@{}c@{}}Read\\Max\end{tabular} & \begin{tabular}[c]{@{}c@{}}Read\\Avg.\end{tabular} & \begin{tabular}[c]{@{}c@{}}Write\\Min\end{tabular} & \begin{tabular}[c]{@{}c@{}}Write\\Max\end{tabular} & \begin{tabular}[c]{@{}c@{}}Write\\Avg.\end{tabular} \\
		\hline
		\texttt{mcause}   & 1345 & 1818 & 1361.16 & 1345 & 1848 & 1358.05 \\
		\texttt{medeleg}  & 1320 & 1924 & 1457.18 & 1324 & 1978 & 1465.66 \\
		\texttt{mepc}     & 1326 & 2000 & 1361.76 & 1328 & 1804 & 1378.52 \\
		\texttt{mhartid}  & 1069 & 1699 & 1075.00 & N/A  &  N/A &    N/A \\
		\texttt{mideleg}  & 1338 & 1848 & 1360.93 & 1342 & 1776 & 1358.95 \\
		\texttt{mie}      & 1096 & 1590 & 1107.11 & 1092 & 1722 & 1105.41 \\
		\texttt{mip}      & 1344 & 1844 & 1479.20 & 1344 & 2018 & 1478.84 \\
		\texttt{misa}     & 1332 & 2006 & 1403.65 & 1340 & 1870 & 1387.85 \\
		\texttt{mscratch} & 1350 & 1827 & 1382.78 & 1348 & 1930 & 1387.53 \\
		\texttt{mstatus}  & 1628 & 2126 & 1671.50 & 1101 & 1398 & 1110.49 \\
		\texttt{mtval}    & 1330 & 2016 & 1389.22 & 1339 & 1884 & 1387.13 \\
		\texttt{mtvec}    & 1586 & 2164 & 1772.32 & 1588 & 2356 & 1775.89 \\
		\hline
	\end{tabular}
\end{table}

Because \texttt{mhartid} is a read-only CSR, the write-latency entries are marked as N/A.

Table \ref{tab:csr_overhead_all} shows that accesses to the M-mode CSRs implemented by the M-mode Emulator have average latencies of approximately 1000 to 1800 cycles.
Table \ref{tab:csr_overhead_compare} shows the overhead of representative CSR accesses relative to an M-mode bare-metal baseline.
The overhead is calculated by subtracting the average latency on M-mode bare metal from the average latency on VioletVM.
For the representative CSR accesses, the M-mode Emulator introduces an additional overhead of approximately 1088 to 1658 cycles.
This overhead occurs because M-mode CSR accesses by the guest OS are trapped by Violet and processed in software by the M-mode Emulator.

\begin{table}[t]
	\centering
	\caption{Representative M-mode CSR access latency on M-mode bare metal and VioletVM (cycles).}
	\label{tab:csr_overhead_compare}
	\setlength{\tabcolsep}{4pt}
	\begin{tabular}{lllrrrr}
		\hline
		\hline
		CSR & Op. & Env. & Min & Max & Avg. & \begin{tabular}[c]{@{}c@{}}Over-\\head\end{tabular} \\
		\hline
		\texttt{mcause} & Read & \begin{tabular}[c]{@{}l@{}}M-mode\\bare metal\end{tabular} & 14 & 14 & 14.00 & 0.00 \\
		\texttt{mcause} & Read & VioletVM          & 1345 & 1818 & 1361.16 & 1347.16 \\
		\texttt{mstatus} & Read & \begin{tabular}[c]{@{}l@{}}M-mode\\bare metal\end{tabular} & 14 & 14 & 14.00 & 0.00 \\
		\texttt{mstatus} & Read & VioletVM          & 1628 & 2126 & 1671.50 & 1657.50 \\
		\hline
		\texttt{mcause} & Write & \begin{tabular}[c]{@{}l@{}}M-mode\\bare metal\end{tabular} & 14 & 14 & 14.00 & 0.00 \\
		\texttt{mcause} & Write & VioletVM          & 1345 & 1848 & 1358.05 & 1344.05 \\
		\texttt{mstatus} & Write & \begin{tabular}[c]{@{}l@{}}M-mode\\bare metal\end{tabular} & 22 & 22 & 22.00 & 0.00 \\
		\texttt{mstatus} & Write & VioletVM          & 1101 & 1398 & 1110.49 & 1088.49 \\
		\hline
	\end{tabular}
\end{table}

\begin{figure}[t]
	\centering
	\includegraphics[width=\columnwidth,keepaspectratio]{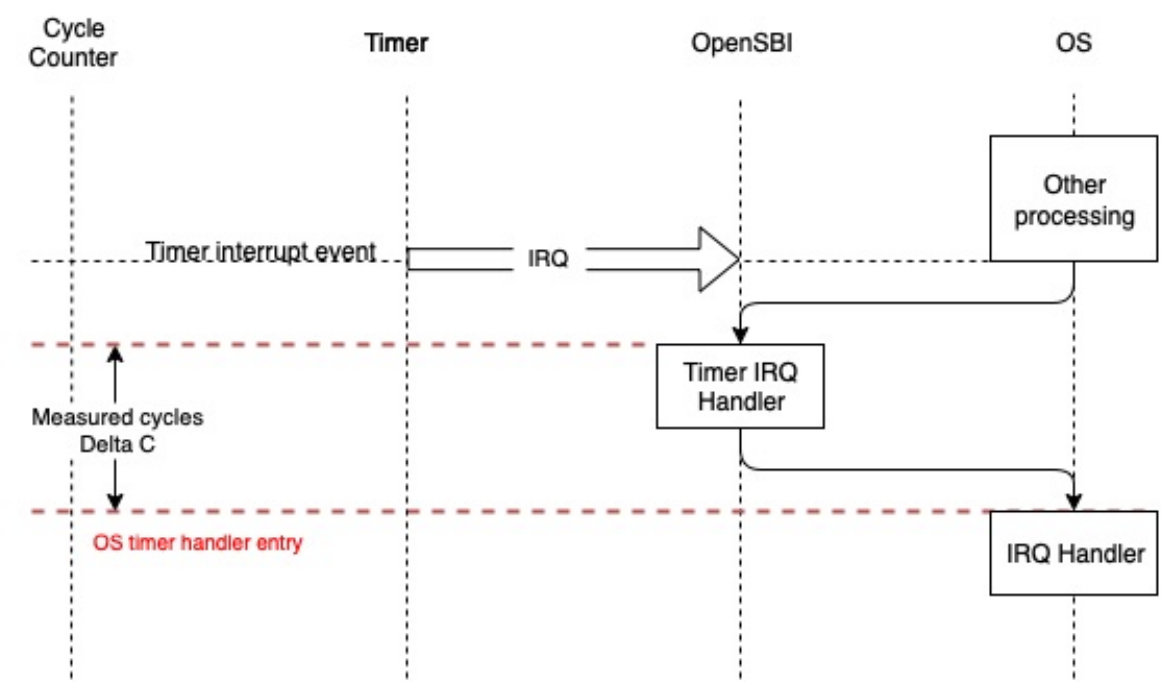}
	\caption{Cycle-count measurement interval for timer interrupt latency on bare metal.}
	\label{fig:timer_latency_measurement_bare}
\end{figure}

\begin{figure}[t]
	\centering
	\includegraphics[width=\columnwidth,keepaspectratio]{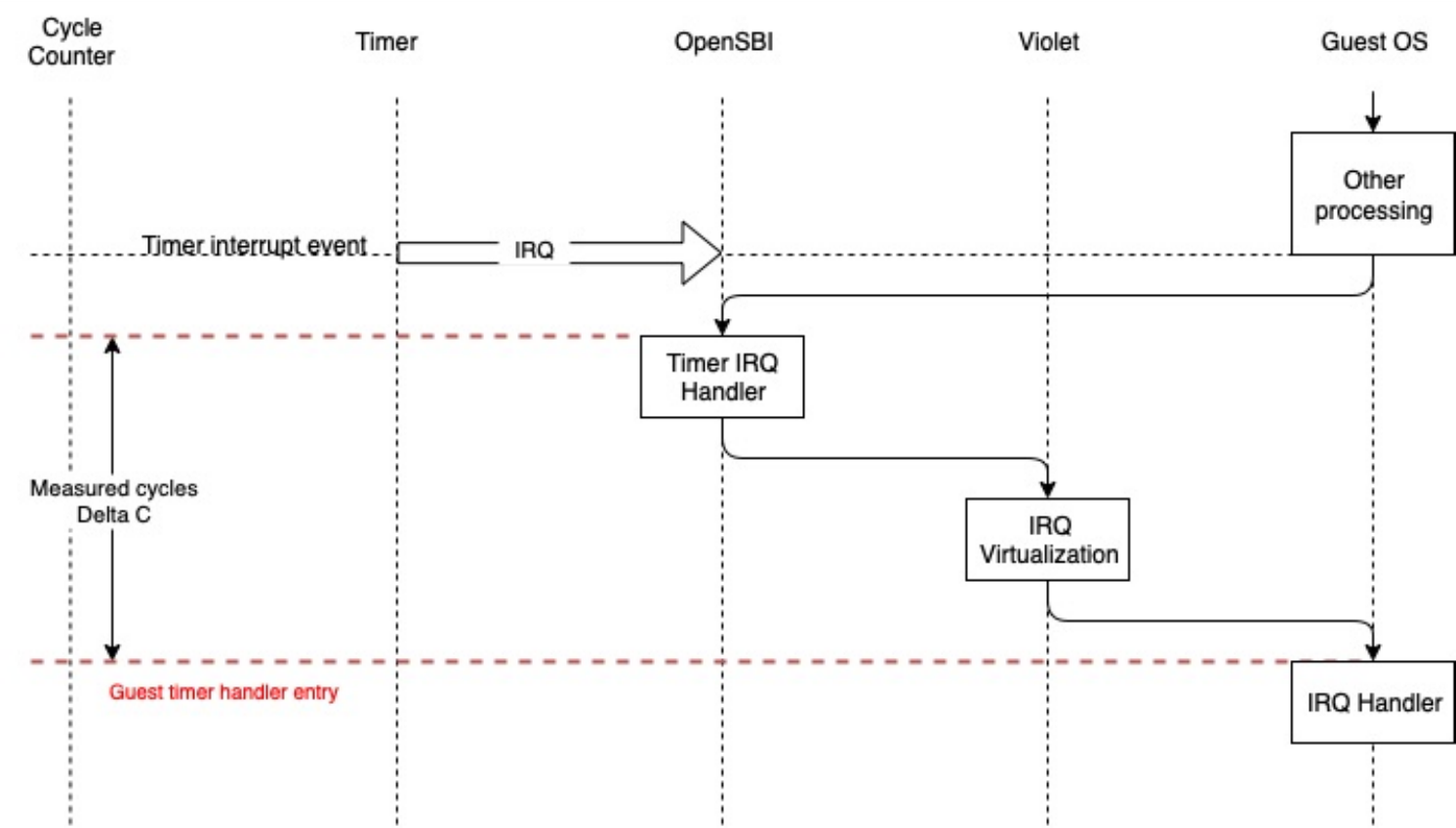}
	\caption{Cycle-count measurement interval for timer interrupt latency on VioletVM.}
	\label{fig:timer_latency_measurement_violet}
\end{figure}

\subsubsection{Timer Interrupt Latency}

Timer interrupt latency was measured to evaluate the effect on RTOS interrupt response.
This measurement records the time from the occurrence of a timer interrupt to the arrival at the RTOS timer interrupt handler.
Figure \ref{fig:timer_latency_measurement_bare} shows the measurement interval for timer interrupt latency on bare metal.
On bare metal, the latency was calculated from the cycle count between the timer interrupt event and entry to the OS timer interrupt handler.
Figure \ref{fig:timer_latency_measurement_violet} shows the measurement interval for timer interrupt latency on VioletVM.
On VioletVM, the endpoint of the same interval was entry to the guest OS timer interrupt handler, so the measurement includes the interrupt virtualization performed before delivery to the guest OS.
The measured cycle count $\Delta C$ was converted to time as $\Delta C / 1.4~\mathrm{GHz}$ using the CPU frequency in this evaluation.
Table \ref{tab:timer_latency} shows the timer interrupt latency in the bare-metal and VioletVM environments in us.

\begin{table}[t]
	\centering
	\caption{Timer interrupt latency (us).}
	\label{tab:timer_latency}
	\begin{tabular}{lrrrr}
		\hline
		\hline
		Environment & Min & Max & Avg. & Median \\
		\hline
		Bare-metal & 0.16 & 0.52 & 0.16 & 0.16 \\
		VioletVM   & 0.66 & 1.31 & 1.05 & 1.03 \\
		\hline
	\end{tabular}
\end{table}

The maximum observed timer interrupt latency was 0.52 us on bare metal and 1.31 us on VioletVM.
Thus, the difference between the maximum observed latencies in this measurement was 0.79 us.

\subsubsection{Thread-Metric Cooperative Scheduling}

The Thread-Metric Testsuite Cooperative Scheduling Test was then executed to evaluate RTOS context-switch performance.
In this test, five tasks repeatedly yield the CPU and switch for 30 seconds.
The score is the total number of task executions, and a higher score indicates a shorter context-switch time.
Because the Thread-Metric Cooperative Scheduling Test does not directly output the context-switch time, the switch time was estimated using (\ref{eq:thread_metric_switch_time}), where $T_{\mathrm{test}}$ is the test duration and $S_{\mathrm{score}}$ is the score.
The overhead was calculated by subtracting the estimated switch time on bare metal from that on VioletVM.
Table \ref{tab:thread_metric} shows the results in the bare-metal and VioletVM environments.

\begin{equation}
	T_{\mathrm{switch}} \approx \frac{T_{\mathrm{test}}}{S_{\mathrm{score}}}
	\label{eq:thread_metric_switch_time}
\end{equation}

\begin{table}[t]
	\centering
	\caption{Thread-Metric cooperative scheduling performance.}
	\label{tab:thread_metric}
	\resizebox{\linewidth}{!}{%
	\begin{tabular}{lrrr}
		\hline
		\hline
		Environment & \multicolumn{1}{c}{Score} & \multicolumn{1}{c}{\begin{tabular}[c]{@{}l@{}}Est. \\ Switch Time (us)\end{tabular}} & \multicolumn{1}{c}{\begin{tabular}[c]{@{}l@{}}Est. \\ Overhead (us)\end{tabular}} \\
		\hline
		Bare-metal & 186,968,445 &  0.16 &  0.00 \\
		VioletVM   &     966,529 & 31.04 & 30.88 \\
		\hline
	\end{tabular}%
	}
\end{table}

The estimated context-switch time was 0.16 us in the bare-metal environment and 31.04 us on VioletVM.
This corresponds to an estimated overhead of 30.88 us compared with bare metal.

The functional evaluation showed that on real RISC-V hardware, Violet's M-mode Emulator can appropriately provide an M-mode execution environment, and Linux and FreeRTOS can run as guest OSs simultaneously.
The performance evaluation quantified the overhead introduced by the M-mode Emulator on M-mode CSR accesses, interrupt response, and context switching.
This overhead originates from software trap-and-emulate-based M-mode emulation and is the cost required to run existing M-mode RTOS binaries without modification.

\section{Related Work} \label{section:related}

Miralis\cite{10.1145/3731569.3764826} is similar to Violet in that it virtualizes M-mode in RISC-V. Still, its purpose is to protect the OS and enclaves from malicious firmware by running M-mode firmware in a virtualized environment. Therefore, it does not mention the co-execution of multiple OSs through the virtualization of RTOS and GPOS. Also, while Violet combines hardware virtualization by the Hypervisor Extension with software virtualization, Miralis virtualizes M-mode without using the Hypervisor Extension.

BaoHypervisor\cite{bao,10.1109/TVLSI.2023.3302837,bao_repo} is a hypervisor aimed at strong isolation between VMs and guaranteeing real-time performance. It is also possible to run FreeRTOS as a guest OS on RISC-V, but it uses a modified version of FreeRTOS\cite{freertos_over_bao} that can run in S-mode, and full virtualization has not been achieved.

HSP-V\cite{10528286} enables the execution of multiple OSs on RISC-V by allocating resources to guests through static partitioning. It does not perform virtualization and runs on RISC-V cores where the Hypervisor Extension is not implemented. Although it mentions running RTOS, since HSP-V runs in M-mode and executes guest OSs in S-mode, it is considered to use a modified version of FreeRTOS that runs in S-mode, similar to BaoHypervisor.

RVirt\cite{rvirt} is a hypervisor that enables guest operating systems to run on a virtual machine through trap-and-emulate, similar to Violet's M-mode Emulator. Therefore, it operates on RISC-V cores that do not implement the Hypervisor Extension. However, it does not target M-mode virtualization for RTOS running.

Xvisor\cite{xvisor} is an embedded hypervisor that supports many architectures. It supports full virtualization and can run guests without modification. While complex configurations such as running multiple OSs simultaneously and communicating between them are possible on ARM and x86\_64, it does not mention running RTOS on the RISC-V architecture.

There are other hypervisors with implementations for RISC-V\cite{kvm,10.1145/1165389.945462,bhyve}, but they target GPOS running in S-mode and cannot run RTOS running in M-mode.

\section{Conclusion} \label{section:conclusion}

In this paper, Violet, an embedded hypervisor for RISC-V, is proposed. By emulating M-mode with the M-mode Emulator, Violet enables the execution of an RTOS in a VM running RISC-V. This allows existing RTOSs to run on a VM in RISC-V and coexist with GPOS such as Linux. Simultaneous execution of FreeRTOS and Linux on actual RISC-V hardware was demonstrated. In addition, RISC-V architecture tests showed that the M-mode Emulator can provide an M-mode execution environment consistent with the RISC-V specification.

In addition, the performance overhead of the M-mode Emulator was evaluated, and its effects on M-mode CSR access, timer interrupt latency, and context switching were quantified.
Future work includes considering measures to reduce the overhead introduced by the M-mode Emulator, such as offloading to hardware if necessary.
Strict interference isolation through shared hardware resources, such as shared caches and memory bandwidth, will also be considered when hardware implementing RISC-V QoS specifications becomes available, such as the Ssqosid extension for QoS identifiers and CBQRI for capacity and bandwidth allocation and monitoring\cite{riscv-ssqosid,riscv-cbqri}.

\bibliographystyle{IEEEtran}
\bibliography{references}

\EOD

\end{document}